\documentclass[pra,twocolumn,smath,amssymb,superscriptaddress]{revtex4-1}

\usepackage{epsfig,amsmath}
\usepackage{subfigure}
\usepackage{graphicx}
\usepackage{dcolumn}
\usepackage{stmaryrd}
\usepackage{mathrsfs}
\usepackage{pifont}
\usepackage{amsthm}
\usepackage{amssymb}
\usepackage{bm}
\usepackage{latexsym}
\usepackage[colorlinks=true,linkcolor=blue,citecolor=blue]{hyperref}
\usepackage{color}
\usepackage{epstopdf}

\newcommand{\non}{\nonumber}

\begin{document}

\title{Coherent noise cancellation in optomechanical system with double optical modes}

\author{Jiashun Yan}
\author{Jun Jing}
\email{Email address: jingjun@zju.edu.cn}
\affiliation{Zhejiang Province Key Laboratory of Quantum Technology and Device, Department of Physics, Zhejiang University, Hangzhou 310027, Zhejiang, China}

\begin{abstract}
The coherent quantum noise cancellation (CQNC) strategy has been performed in the single-mode optomechanical systems to promote an ultra-sensitive metrology protocol to break the standard quantum limit. The key idea of CQNC is that the backaction noises arising from radiation pressure and driving can be offset by coupling the optical mode to a near-resonant ancillary mode. In this work, a continuous weak-force sensing under CQNC is developed in a double-mode optomechanical system consisted of two optical modes with distinct frequencies and a mechanical mode. In particular, under the asymmetrical treatment by driving the higher-frequency optical mode, probing the lower-frequency one, and coupling the probe mode to the ancillary mode, our configuration can be used to resemble the conventional CQNC sensing. It is more important to find that the current CQNC strategy simultaneously stabilizes the double-mode system with respect to both the constrained driving power (the Routh-Hurwitz criterion) and the effective positive mechanical damping (the stable optical-spring condition). Moreover, through exploiting the coupling between the probe mode and the ancillary mode under this nontrivial extension of the CQNC strategy (from the single-mode version to the double-mode one), the rotating-wave term and the counter-rotating term are found to be responsible to the system stability and the noise cancellation, respectively. In realistic situations, our scheme can be practiced in a tripartite optomechanical setup with a membrane in the middle and a twisted-cavity-based weak-torque detector.
\end{abstract}

\maketitle

\section{Introduction}\label{intro}

At the intersection of nanophysics and quantum optics, quantum optomechanics exploits the interactions between the electro-magnetic radiation and the mechanical-oscillator motion. In the Fabry-P\'erot cavity typically used in the cavity optomechanics, the light field exerts a radiation pressure on a vibrating mirror~\cite{cavityOpto}. Myriad applications on the radiation pressure have been realized in the cavity optomechanics, such as cooling the motion of oscillators to their ground states~\cite{cooling1,cooling2,cooling3,cooling4,cooling5}, demonstrating the quantum-to-classical transitions~\cite{optoEntangle}, and controlling the photon transport~\cite{transport}. Among these quantum science and technologies, the ultra-sensitive measurement or detection about the weak force on the mechanical oscillator are under intensive investigations~\cite{sensing1,sensing2,sensing3,sensing4} as a significant branch of quantum metrology~\cite{QuantumSensing,QuantumMetrology}.

Optomechanical systems are also well-known for providing an efficient way of converting the quantum information at MHz frequencies (mechanical motion) up to optical frequencies~\cite{CavityOptomechanicsBooks}. They are subject to hybrid noises from various sources or environmental degrees of freedom. The noises induce errors to sensing or measurement with the mechanical motion, since they cannot be distinguished from the signal and always are undetermined to observers. The weak-force sensing by optomechanical systems then desires a high susceptibility to magnify the external force and an effective suppression over the measurement noise, as the noise might be synchronously amplified. The measurement noise generally comprises the shot noise and the backaction noise~\cite{IntroNoise,Deamplification,RadiaPressureFluctuations}. The shot noise is induced by photon number-phase uncertainty limiting the precision in the interferometric experiments, such as the laser interferometer gravitational-wave observatory~\cite{LIGO}, which decreases with the pumping power. While the backaction noise results from the fluctuations in the radiation-pressure of the optical mode, which increases with the pumping power and was observed for the first time in Refs.~\cite{BAnoise1,BAnoise2}. Then the trade-off between these two noises leads to a lower bound for the detection sensitivity, i.e., the standard quantum limit (SQL)~\cite{IntroNoise,Clerk2008}.

Various methods for the optomechanical force-sensor to break through SQL have been proposed to suppress the backaction noise, such as the frequency-dependent squeezing of the input light~\cite{squeezing2,squeezing3,interferometer,nondemolitionInter}, the variational measurement~\cite{nondemolitionInter,variation1,variation2}, and the application of the dual mechanical resonators~\cite{dual1,dual2} or an optical spring to modify the mechanical response function~\cite{spring1,spring2}. Compared to these backaction-evading techniques, a quantum protocol named coherent quantum noise cancellation (CQNC) recently suggested by Tsang and Caves~\cite{CQNC1} could interfere destructively with the backaction noise by the antinoise path upon coupling to a deliberately designed ancillary mode. In this work, the CQNC strategy as a judicious protocol to reduce the measurement noise is developed from a single-optical-mode optomechanical system~\cite{CQNC2} to a double-optical-mode one. The multi-mode optomechanical systems~\cite{twomodeEntangle,nonreciprocity,blockade} claim to have an ultra-sensitivity by the generated squeezed states due to the interaction between the cavity photons and the mechanical oscillator~\cite{squeezing1} than the single-mode system. While in regard to the CQNC scheme, the multi-mode system might import additional measurement noises raised by the detuning between the system mode and the ancillary mode. Driving the high-frequency mode and detecting the low-frequency one is found to be a nontrivial and crucial prerequisite to realize the noise-suppression in the current scheme. As a result, the double-optical-mode system can be reduced to an effective single-mode one and the driven-mode fluctuation becomes separable in the whole dynamics after the linearization process. In the mean time, we find that the full system-stability is promoted in the presence of the CQNC control. 

Our protocol presents with an avoided normal-mode splitting induced by the strong coupling between the optical modes and the mechanical mode to show the ultra-sensitivity around the mechanical-mode frequency. When the ancillary optical mode is near-resonant with the probe mode, the desired noise cancellation is followed by building an effective coherent channel to compensate the backaction noise. Our scheme can be performed in both a conventional ``membrane-in-the-middle'' setup~\cite{three-mode,MultipleMembrane,squeezing1,StrongDispersiveCoupling} and a novel twisting optomechanical cavity consisting of a torsional mechanical oscillator~\cite{twisting}.

The rest of this work is structured as following. In Sec.~\ref{ModelAndHamiltonian}, we introduce and analyse the free Hamiltonian for the double-mode optomechanical system with two optical modes of distinct frequencies under two asymmetrical configurations in regard to driving and probing. The choice of driving the high-frequency mode and probing the low-frequency mode for breaking through SQL in quantum metrology is supported by the Routh-Hurwitz criterion for the system stability, which places a limit on the driving power. In Sec.~\ref{DetectionNoCQNC}, the susceptibility of the mechanical oscillator and the standard quantum limit are evaluated to show the sensing performance of the free system. Then the normal mode splitting is discussed as a hallmark of the strong coupling. In Sec.~\ref{CQNC}, the microscopic mechanism of the measurement-noise cancellation by the CQNC idea is illustrated under a balanced coupling between the probe and the ancillary modes. A remarkably improved sensitivity for the weak-force metrology is demonstrated to break through the SQL and the system-stability condition in both the Routh-Hurwitz criterion and the stable optical spring are analysed in details. Then in Sec.~\ref{unbalance}, we discuss further the effect on CQNC-sensing and system-stability from the imbalanced coupling about the rotating-wave term and the counter-rotating term. In Sec.~\ref{PhyRealize}, two optomechanical setups are proposed to realize the initial Hamiltonian of our model available for control. We summarize the whole work in Sec.~\ref{conclusion}.

\section{Model and Hamiltonian}\label{ModelAndHamiltonian}

We start with a general optomechanical system consisting of two optical modes with distinct frequencies and a mechanical mode. With one of the optical modes under driving, the system Hamiltonian can be written as ($\hbar\equiv1$)
\begin{eqnarray}\non
H&=&\omega_aa^\dagger a+\omega_b b^\dagger b+\frac{\omega_m}{2}(x^2+p^2)+gx(a^\dagger b+ab^\dagger)\\
\label{Hamiltonian}		&+&iE\left(a^\dagger e^{-i\omega_dt}-ae^{i\omega_dt}\right),
\end{eqnarray}
where $a$, $b$ ($a^\dagger, b^\dagger$) are respectively the annihilation (creation) operators of the high-frequency and the low-frequency optical modes, and $\omega_a$ and $\omega_b$ are their frequencies with $\omega_a>\omega_b$. $x\equiv x_m/x_{\rm ZPF}$ and $p\equiv p_mx_{\rm ZPF}$ are respectively the dimensionless position and momentum operators of the mechanical oscillator. Here $x_{\rm ZPF}\equiv1/\sqrt{m\omega_m}$ is the zero point fluctuation with $\omega_m$ the frequency and $m$ the mechanical-oscillator mass. $g$ is the coupling strength between the optical modes and the mechanical mode and this nonlinear optomechanical interaction can be realized by a double-side radiation-pressure or a permittivity tensor modulation (The details are left to Sec.~\ref{PhyRealize}). $E\equiv\sqrt{P_{\rm in}\kappa_a/\omega_d}$ is the driving strength determined by the driving power $P_{\rm in}$, the cavity damping coefficient $\kappa_a$ and the driving-laser frequency $\omega_d$.

The optomechanical system is supposed to be subject to a Markovian environment through damping loss. Then in the rotating frame with respect to $H_d=\omega_d(a^\dagger a+b^\dagger b)$, a set of Heisenberg-Langevin equations can be obtained through the input-output theory:
\begin{equation}\label{OringalLongevin}
	\begin{aligned}
		&\dot a=-i\Delta_aa-ig x b-\kappa_a a+\sqrt{2\kappa_a}a_{\rm in}+E,\\
		&\dot b=-i\Delta_bb-ig x a-\kappa_b b+\sqrt{2\kappa_b}b_{\rm in},\\
		&\dot x=\omega_m p,\\
		&\dot p=-\omega_m x-g(a^\dagger b+ab^\dagger)-\gamma_m p+F_{\rm in}.
	\end{aligned}
\end{equation}
Here $\Delta_{a,b}\equiv\omega_{a,b}-\omega_d$ is the detuning between the optical mode $a$ ($b$) and the driving laser. $\kappa_i$ ($i=a,b$) and $\gamma_m$ are respectively the relaxation rates of the optical and mechanical modes. $a_{\rm in}$ and $b_{\rm in}$ account for the noise operators associated with the respective input fields. Due to the fluctuation-dissipation theorem, the autocorrelation functions of the vacuum noises for the optical modes satisfy $\langle a_{\rm in}(t)a^\dagger_{\rm in}(\tau)\rangle=\langle b_{\rm in}(t)b_{\rm in}^\dagger(\tau)\rangle=\delta(t-\tau)$. $F_{\rm in}$ consists of the to-be-determined external force $F_{\rm ext}$ acting on the oscillator and the Brownian noise of the oscillator $\xi$ satisfying $\langle\xi(t)\xi(\tau)\rangle\approx2\gamma_m n_{\rm th}\delta(t-\tau)$, where $n_{\rm th}\equiv1/(e^{\omega_m/T}-1)$ denotes the average population of the mechanical oscillator ($k_B\equiv1$). The thermal Brownian noise is assumed to be overwhelmed by the backaction noise~\cite{CQNC2} or can be suppressed by precooling of the mechanical mode~\cite{precooling} before CQNC sensing.

The linearization process by decomposing the operators into the expectation-value part (time-independent) and the fluctuating part (time-dependent) is valid under a sufficiently large amplification by a strong pumping or driving. Under this condition, one can write $O=\langle O\rangle+\delta O$, where $O$ is an arbitrary operator and $\langle O\rangle$ is the expectation value with respect to the steady-state of the system. Inserting the decomposed expressions into Eq.~(\ref{OringalLongevin}), it is found that $\langle a\rangle=\alpha=E/(i\Delta_a+\kappa_a)$, $\langle b\rangle=0$, $\langle x\rangle=0$, and $\langle p\rangle=0$. Omitting the quadratic terms $\delta x\delta b$, $\delta x\delta a$, $\delta a^\dagger\delta b$, and $\delta a\delta b^\dagger$, and reexpressing the fluctuation variables $\delta O\to O$, one can have the linearized quantum Heisenberg-Langevin equations as following (up to a phase modulation over the operator $b$):
\begin{equation}\label{linearLanggevin}
	\begin{aligned}
		&\dot a=-i\Delta_aa-\kappa_a a+\sqrt{2\kappa_a}a_{\rm in}, \\
		&\dot b=-i\Delta_bb-i G x-\kappa_b b+\sqrt{2\kappa_b}b_{\rm in}, \\
		&\dot x=\omega_m p, \\
		&\dot p=-\omega_m x-G(b^\dagger+b)-\gamma_m p+F_{\rm in},
	\end{aligned}
\end{equation}
where $G\equiv|\alpha|g$. Note now the dynamics and the noise field of the driven mode-$a$ are decoupled from those of the probe mode-$b$ and the mechanical oscillator. Nevertheless the coefficient $|\alpha|$ indicates a significant amplification on the effective coupling between the probe mode and the mechanical mode.

It is interesting to make an argument here about the choice of the driven mode and the probe mode in the initial Hamiltonian~(\ref{Hamiltonian}) before further discussion. First, the choice of simultaneously driving both optical modes can be ruled out. In that scenario, the expectation value $\langle b\rangle$ at the steady state will be nonzero, so that after the linearization process the mode $a$ can {\em not} be decoupled from the whole dynamics hindering the ensued quantum metrology. Second, if one drives the low-frequency optical mode-$b$ and probes the high-frequency one, i.e., modifies the driving term in Eq.~(\ref{Hamiltonian}) by $a\to b$, then through a similar derivation one can arrive at
\begin{equation}\label{linearLanggevinRedDetuning}
	\begin{aligned}
		&\dot b=-i\Delta_bb-\kappa_b b+\sqrt{2\kappa_b}b_{\rm in},\\
		&\dot a=-i\Delta_aa-i G' x-\kappa_a a+\sqrt{2\kappa_a}a_{\rm in},\\
		&\dot x=\omega_m p,\\
		&\dot p=-\omega_m x-G'(a^\dagger+a)-\gamma_m p+F_{\rm in},
	\end{aligned}
\end{equation}
where $G'=|\langle b\rangle|g$. An intuitive insight tells no essential difference between Eqs.~(\ref{linearLanggevin}) and (\ref{linearLanggevinRedDetuning}). Both of them resemble an effective single-mode optomechanical system. However, to solely drive one of these two modes with distinct frequencies, $\omega_d$ has to be near-resonant with $\omega_a$ to attain Eq.~(\ref{linearLanggevin}) or be near-resonant with $\omega_b$ to attain Eq.~(\ref{linearLanggevinRedDetuning}). Then the coefficients $\Delta_b<0$ in Eq.~(\ref{linearLanggevin}) and $\Delta_a>0$ in Eq.~(\ref{linearLanggevinRedDetuning}), correspond respectively to the blue-detuning and the red-detuning cases in practice. The linearized Heisenberg-Langevin equations~(\ref{linearLanggevin}) and (\ref{linearLanggevinRedDetuning}) are therefore {\em not} symmetrical to each other, even upon $a\leftrightarrow b$. These two asymmetrical configurations will manifest dramatically different results for the weak-force detection in terms of system-stability, normal-mode splitting and CQNC sensing (The detailed analysis can be found in the following sections). In short, we find that driving the high-frequency mode and probing the low-frequency mode are available to break through SQL in quantum metrology nearby the mechanical frequency. This nontrivial choice justifies our Hamiltonian in Eq.~(\ref{Hamiltonian}) as well as the dynamics by Eq.~(\ref{linearLanggevin}) in the double-mode version for the CQNC strategy.

From Eq.~(\ref{linearLanggevin}), a large coupling-strength $G$ determined by the initial coupling-strength $g$ and the average photon number in the cavity $|\alpha|^2$ is demanded for CQNC sensing~\cite{CQNC2}. The Routh-Hurwitz stability criterion, however, restricts the value of $|\alpha|$. In particular, the real part of all roots of the characteristic polynomial for the system has to be negative~\cite{RHcriterion} to ensure the stability of the linear system. For our optomechanical system, the Routh-Hurwitz criterion yields
\begin{subequations}
\begin{equation}\label{AlphaBlueDetuing}
|\alpha|^2>-\frac{(\Delta_b^2+\kappa_b^2)\omega_m}{2|\Delta_b| g^2},
\end{equation}
under the blue-detuning situation $\Delta_b<0$ and
\begin{equation}\label{AlphaRedDetunig}
|\alpha|^2<\frac{(\Delta_b^2+\kappa_b^2)\omega_m}{2|\Delta_b| g^2},
\end{equation}
\end{subequations}
under the red-detuning situation $\Delta_b>0$ [make $a\rightarrow b$ in Eq.~(\ref{linearLanggevinRedDetuning})], respectively. The detailed calculation is provided in appendix~\ref{StabilityCondition}. Clearly even with the free Hamiltonian, the average photon number $|\alpha|^2$ determined by the driving power is under the magnitude restriction when driving the low-frequency optical mode while probing the high-frequency one, i.e., in the red-detuning situation $\Delta_b>0$. In contrast, the system stability is always guaranteed in the blue-detuning situation $\Delta_b<0$, since Eq.~(\ref{AlphaBlueDetuing}) always holds.

\section{Weak Force Detection without CQNC}\label{DetectionNoCQNC}

\subsection{Mechanical susceptibility and standard quantum limit}\label{SusceptibilityAndSQL}

For quantum metrology, it is convenient to transform the time evolution of system into the frequency domain to analyse the linear response in the noise spectrum of the mechanical oscillation to the external force~\cite{cavityOpto}. By the Fourier transformation $O(\omega)\equiv\frac{1}{2\pi}\int dtO(t)e^{i\omega t}$ for all the operators, the dynamics of the system (dropping the driven mode-$a$) in Eq.~(\ref{linearLanggevin}) can then be displayed in the frequency space by
\begin{equation}\label{LongevinFrequency}
	\begin{aligned}
		& -i\omega x=\omega_m p,\\
		&(\gamma_m-i\omega) p=-\omega_m x-\sqrt{2}G x_b+F_{\rm in},\\
		&(\kappa_b-i\omega) x_b=-\kappa_b x_b+\Delta_bp_b+\sqrt{2\kappa_b}x_{\rm in}^b,\\
		&(\kappa_b-i\omega) p_b=-\Delta_b x_b-\sqrt{2}G x+\sqrt{2\kappa_b}p_{\rm in}^b,
	\end{aligned}
\end{equation}
where the quadratures are $x_b\equiv(b+b^\dagger)/\sqrt{2}$, $p_b\equiv(b-b^\dagger)/\sqrt{2}i$ and the relevant noise operators are $x_{\rm in}^b\equiv(b_{\rm in}+b_{\rm in}^\dagger)/\sqrt{2}$, $p_{\rm in}^b\equiv(b_{\rm in}-b_{\rm in}^\dagger)/\sqrt{2}i$. Solving the linear equations~(\ref{LongevinFrequency}), one could find $x(\omega)$ as a function of variables $F_{\rm in}$, $x_{\rm in}^b$ and $p_{\rm in}^b$:
\begin{eqnarray}\nonumber
x(\omega)&=&\chi(\omega)\Biggl\{F_{\rm in}(\omega)-\frac{2\sqrt{\kappa_b}G}{(\kappa_b-i\omega)^2+\Delta_b^2}\\
&\times&\left[(\kappa_b-i\omega)x_{\rm in}^b(\omega)+\Delta_bp_{\rm in}^b(\omega)\right]\Biggr\}, \label{xomega}
\end{eqnarray}
where $\chi(\omega)$ is defined as the susceptibility of the mechanical oscillation:
\begin{equation}\label{susceptibility}	\chi(\omega)=\left[\frac{\omega_m^2-i\omega\gamma_m-\omega^2}{\omega_m}-\frac{2G^2\Delta_b}
{(\kappa_b-i\omega)^2+\Delta_b^2}\right]^{-1}.
\end{equation}
In the linear-response function described by Eq.~(\ref{xomega}), the optomechanical system can be viewed as a linear amplifier for the input fields from the probe mode $b$ and the mechanical oscillator. The real and imaginary parts of the susceptibility in Eq.~(\ref{susceptibility}) imply respectively the dissipation rate (proportional to $G^2$) and the mechanical-frequency shift due to the inner-couplings of the optomechanical system.

The phase shift of the transmitted or the reflected light of the optical-mode $b$ allows an indirect measurement over the displacement of the mechanical-oscillator under the external force. This measurement is often performed with a homodyne detector, in which the signal is brought to interfere with a local oscillator as a phase reference. Based on the input-output theory, we have
\begin{equation}\label{InOut}
	\begin{aligned}
		& x_{\rm out}^b(\omega)=\sqrt{2\kappa_b}x_b(\omega)-x_{\rm in}^b(\omega), \\
		& p_{\rm out}^b(\omega)=\sqrt{2\kappa_b}p_b(\omega)-p_{\rm in}^b(\omega).
	\end{aligned}
\end{equation}
The output field will carry information about the inner field, then the force acting on the mechanical oscillator could be estimated by the continuous homodyne measurement over the quadratures of the output signal
\begin{eqnarray}\label{M}
M(\omega)&=&\sin\varphi x_{\rm out}^b(\omega)+\cos\varphi p_{\rm out}^b(\omega)\\ \nonumber
	&=&\chi_F(\omega)F_{\rm in}(\omega)+\chi_x(\omega)x_{\rm in}^b(\omega)+\chi_p(\omega)p_{\rm in}^b(\omega),
\end{eqnarray}
where
\begin{subequations}
\begin{equation}\label{XF}	
\chi_F(\omega)=\frac{-2\sqrt{\kappa_b}G\bigl[\Delta_b\sin\varphi+(\kappa_b-i\omega)
\cos\varphi\bigr]}{(\kappa_b-i\omega)^2+\Delta_b^2}\chi(\omega),
\end{equation}
\begin{equation}\label{XX}
	\begin{aligned}		
&\chi_x(\omega)=\frac{(\kappa_b^2+\omega^2-\Delta_b^2)\sin\varphi-2\kappa_b\Delta_b\cos\varphi}
{(\kappa_b-i\omega)^2+\Delta_b^2} \\ 		
&+\frac{4\kappa_bG^2(\kappa_b-i\omega)\big[\Delta_b\sin\varphi+(\kappa_b-i\omega)\cos\varphi\big]}
{[(\kappa_b-i\omega)^2+\Delta_b^2]^2}\chi(\omega), \\
	\end{aligned}
\end{equation}
\begin{equation}\label{XP}
	\begin{aligned}		
&\chi_p(\omega)=\frac{(\kappa_b^2+\omega^2-\Delta_b^2)\cos\varphi+2\kappa_b\Delta_b\sin\varphi}
{(\kappa_b-i\omega)^2+\Delta_b^2}\\
		&+\frac{4\kappa_b G^2\Delta_b\big[\Delta_b\sin\varphi+(\kappa_b-i\omega)\cos\varphi\big]}
{[(\kappa_b-i\omega)^2+\Delta_b^2]^2}\chi(\omega),
	\end{aligned}
\end{equation}
\end{subequations}
with $\varphi$ the phase of the local oscillator (LO) field modulated by an electro-optical modulator~\cite{homodyne}. For simplicity, $\varphi$ is set as zero in following discussion. So that we only measure the phase quadrature by homodyne detection. $\chi_F(\omega)$ characteristics the amplification of the detection signal. $\chi_x(\omega)$ and $\chi_p(\omega)$ are regarded as the noisy signals in comparison to $\chi_F(\omega)$. To analyse the spectral density of these measurement noises, one can define an effective force noise:
\begin{eqnarray}\nonumber
	F_{N}(\omega)&\equiv&\frac{M(\omega)}{\chi_F(\omega)}-F_{\rm ext}(\omega)  \\  \label{ForceNoise}
   &=& \xi(\omega)+\frac{\chi_x(\omega)}{\chi_F(\omega)}x_{\rm in}^b(\omega)+\frac{\chi_p(\omega)}{\chi_F(\omega)}p_{\rm in}^b(\omega).
\end{eqnarray}
Then the quantum noise spectrum $S(\omega)$ can be obtained by~\cite{correlation}
\begin{subequations}\label{NoiseSpectrumDefinition}
	 \begin{align}
		& S(\omega)=\frac{1}{2}\left[S_{FF}(\omega)+S_{FF}(-\omega)\right], \\	 	
        & S_{FF}(\omega)=\int d\omega'\langle F_{N}(\omega)F_{N}(\omega')\rangle,
	\end{align}
\end{subequations}
where the thermal-noise spectrum is expressed by
\begin{equation}\label{Sth}
	S_{\rm th}(\omega)=\int d\omega'\langle \xi(\omega)\xi(\omega')\rangle\approx 2\gamma_m n_{\rm th}.
\end{equation}
Then due to the fact that the vacuum input noise $b_{\rm in}(t)$ satisfies the $\delta$-correlation function, the noise spectrum without CQNC is expressed by
\begin{eqnarray}\nonumber
S(\omega)&=&S_{\rm th}+\frac{\Delta_b(\Delta_b^2-\omega^2+3\kappa_b^2)\Delta}{2\kappa_b\omega_m(\omega^2+\kappa_b^2)}
+\frac{G^2(\Delta_b^2+4\kappa_b^2)}{2\kappa_b(\omega^2+\kappa_b^2)}\\ \label{NoiseSpectrum}
&+& \frac{\bigl[(\Delta_b^2-\omega^2+\kappa_b^2)^2+4\kappa_b^2\omega^2\bigr]
(\omega^2\gamma_m^2+\Delta^2)}{8G^2\kappa_b\omega^2_m(\omega^2+\kappa_b^2)},
\end{eqnarray}
where $\Delta\equiv\omega^2-\omega_m^2$. The second term on the right-hand side of Eq.~(\ref{NoiseSpectrum}) takes the role of the background noise (independent of the coupling-strength $G$) induced by the detuning between the measured frequency and the mechanical-oscillator frequency. The third term scaling as $G^2$ denotes the backaction noise. The last term scaling as $1/G^2$ is the shot noise or the imprecision noise~\cite{Deamplification,IntroNoise}. The lower bound of the noise spectrum with the optimized value of $G$ in Eq.~(\ref{NoiseSpectrum})
\begin{equation}\label{SQLG}
	\begin{aligned}		G_L=\left\{\frac{\left[(\Delta_b^2-\omega^2+\kappa_b^2)^2+4\kappa_b^2\omega^2\right]
(\omega^2\gamma_m^2+\Delta^2)}{4\Delta_b^2\omega_m^2+16\kappa_b^2\omega_m^2}\right\}^{\frac{1}{4}}
	\end{aligned}
\end{equation}
is the standard quantum limit for continuous force sensing:
\begin{equation}\label{SQL}
	\begin{aligned}
		&S_L(\omega)=S_{\rm th}+\frac{1}{2\kappa_b\omega_m(\omega^2+\kappa_b^2)}		
		\Big[\Delta_b(\Delta_b^2-\omega^2+3\kappa_b^2)\Delta+ \\ 		&\sqrt{(\Delta_b^2+4\kappa_b^2)(\omega^2\gamma_m^2+\Delta^2)}
\sqrt{(\Delta_b^2-\omega^2+\kappa_b^2)^2+4\kappa_b^2\omega^2}\Big].
	\end{aligned}
\end{equation}
It places a limit on the detector sensitivity about the weak signal, which can be certainly broken though via a control strategy down to the quantum level, such as the coherent quantum noise cancellation.

\subsection{Strong coupling and normal-mode splitting}

\begin{figure}[htbp]
\centering
\includegraphics[width=0.9\linewidth]{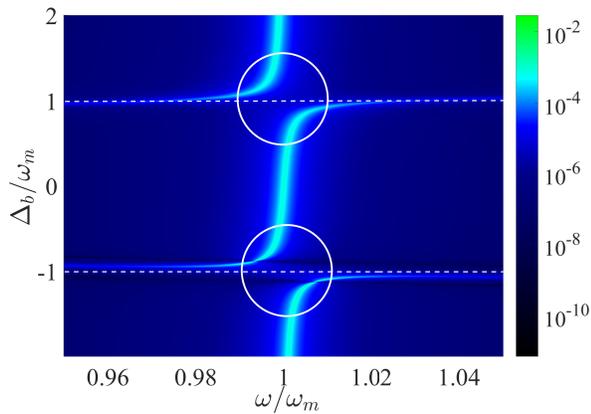}
\caption{(Color online) The absolute value of the imaginary part of the mechanical susceptibility $|{\rm Im}\chi(\omega)|$ as a function of the normalized frequency $\omega/\omega_m$ and the detuning $\Delta_b/\omega_m$. The other parameters are set as $\kappa_b=10^{-2}\omega_m$, $G=4\kappa_b$ and $\gamma_m=1.2\times10^{-3}\omega_m$.}	\label{NMS1}
\end{figure}

The strong coupling between the probe mode and the mechanical mode is a prerequisite for CQNC sensing. It is usually marked by the normal-mode splitting (NMS) or the avoided NMS. In general, NMS occurs in a coupled two-partite system with an energy-exchange interaction larger than the decay rate of the system~\cite{optoNMS}. This section is contributed to analyzing the detection sensitivity through the NMS phenomena under the red-sideband $\Delta_b=\omega_m$ and the blue-sideband $\Delta_b=-\omega_m$ situations, which correspond to the case of driving the low-frequency optical mode and probing the high-frequency optical mode and the opposite case, respectively.

In Fig.~\ref{NMS1}, the absolute value of the imaginary part of the mechanical-mode susceptibility $|{\rm Im}\chi(\omega)|$ is plotted as a function of the normalized frequency $\omega/\omega_m$ and detuning $\Delta_b/\omega_m$. ${\rm Im}\chi(\omega)$ is regarded as the effective dissipation rate for the mechanical oscillator and used to identify the NMS phenomena~\cite{CavityOptomechanicsBooks} in the parametric space. Under the condition $G=4\kappa_b$, the (red and blue) sidebands exhibit a clear bifurcation demonstrating that the (realistic and avoided) mode-splitting and the detection sensitivity will obtain extreme values around these bifurcations. Note if the effective coupling strength is less than the optical dissipation rate $G<\kappa_b$, then the splitting around the two sidebands would not occur any more.

\begin{figure}[htbp]
\centering
\includegraphics[width=0.9\linewidth]{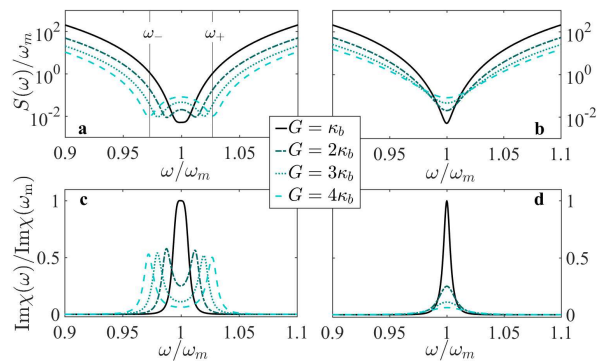}
\caption{(Color online) The normalized noise spectrum $S(\omega)/\omega_m$ as a function of the normalized frequency $\omega/\omega_m$ with various coupling strength $G$ for (a) the red-sideband $\Delta_b=\omega_m$ and (b) the blue-sideband $\Delta_b=-\omega_m$ situations, respectively. The ratio of the imaginary parts of the mechanical susceptibilities $\rm Im\chi(\omega)/\rm Im\chi(\omega_m)$ as a function of $\omega/\omega_m$ for (c) the red-sideband and (d) the blue-sideband situation, respectively. Note here $\chi(\omega_m)$ is the susceptibility with a fixed coupling strength $G=\kappa_b$ for the mechanical frequency $\omega_m$. The dissipation rates of the probe mode and the mechanical mode are respectively $\kappa_b=10^{-2}\omega_m$ and $\gamma_m=1.2\times10^{-3}\omega_m$. The average population for the thermal noise is fixed as $n_{\rm th}=10$~\cite{correlation}. }\label{NMS2}
\end{figure}

The NMS phenomenon can be deliberately observed through the noise spectrum and the susceptibility with various coupling strengths. In the red-sideband situation [see Fig.~\ref{NMS2}(a)(c)], both the noise spectrum and the susceptibility are splitted into a double-valley or a double-peak pattern symmetrical to the mechanical-oscillator frequency when $G>\kappa_b$, where the peak or valley separation becomes even larger with a stronger coupling. In contrast, the avoided NMS pattern is clearly observed in the blue-sideband situation [see Fig.~\ref{NMS2}(b)(d)]. The susceptibility displays a single-valley or a single-weak pattern around $\omega_m$, featuring a larger amplification in magnitude in the near-resonant regime for $\omega_m$ under a stronger coupling strength, that is meaningful to the weak-force sensing.

Both NMS and avoided NMS phenomena can be understood by the linearized Heisenberg-Langevin equations~(\ref{linearLanggevin}) describing the fluctuation dynamics around the steady states of both the mechanical mode and the probe mode. Equation~(\ref{linearLanggevin}) can be obtained alternatively by the effective Hamiltonian:
\begin{equation}\label{Effective_Hamiltonian}
H_{\rm eff}=\Delta_b'b^\dagger b+\omega_m'm^\dagger m+\frac{G}{\sqrt{2}}(m^\dagger +m)(b^\dagger +b),
\end{equation}
where $\Delta_b'\equiv\Delta_b-i\kappa_b$, $\omega_m'\equiv\omega_m-i\gamma_m$ and $m$ ($m^\dagger$) is the annihilation (creation) operator for the mechanical mode, i.e., $x=(m+m^\dagger)/\sqrt{2}, p=(m-m^\dagger)/\sqrt{2}i$. Note this Hamiltonian is non-Hermitian due to the phenomenological dissipation and shares a similar formation as the initial Hamiltonian in Ref.~\cite{optoNMS}. The mechanical mode and the probe mode in this effective Hamiltonian are coupled by a Rabi interaction, which is the sum of the Jaynes-Cummings or rotating-wave interaction leading to the resonant splitting of modes and the counter-rotating terms that do not contribute to the stable dressed states. In the red-detuning situation $\Delta_b>0$, the rotating-wave terms will survive under the long-time average via rotating the Hamiltonian in Eq.~(\ref{Effective_Hamiltonian}) to the interaction picture with respect to $\Delta_bb^\dagger b+\omega_mm^\dagger m$. While in the blue-detuning situation $\Delta_b<0$, however, the counter-rotating terms become dominant.

A further explanation could be made about the normalized frequencies, i.e., the eigenvalues of the Hamiltonian in Eq.~(\ref{Effective_Hamiltonian}). By the Bogoliubov transformation~\cite{stateTransfer}, they read
\begin{eqnarray}\nonumber	\omega_\pm&=&\sqrt{\frac{\Delta_b'^2+\omega_m^2}{2}\pm\frac{1}{2}\sqrt{(\Delta_b'^2-\omega_m'^2)^2+8G^2\omega_m'\Delta_b'}}
\\ &\approx&\sqrt{\omega_m^2\pm\sqrt{2G^2\Delta_b\omega_m}},
\end{eqnarray}
where the approximated solution in the second line is obtained with a large mechanical quality factor $\omega_m\gg\gamma_m$ and a resolved sideband $\omega_m\gg\kappa_b$. With the red-sideband $\Delta_b=\omega_m$, the frequency splitting $\Delta\omega\equiv\omega_+-\omega_-$ is significant in the real part [see the normal-mode splitting in Fig.~\ref{NMS2}(a)(c)], unless $G/\omega_m>1/\sqrt{2}$ leaving the linear system unstable due to Eq.~(\ref{AlphaRedDetunig}). In addition, the splitting $\Delta\omega$ is proportional to the square root of the enhanced coupling strength $\sqrt{G}$. With the blue-sideband $\Delta_b=-\omega_m$, however, $\Delta\omega$ is dominated by an imaginary part and merely affects the modulus of the normal-mode frequencies. It thus gives rise to the avoided NMS in Fig.~\ref{NMS2}(b)(d).

Comparing Fig.~\ref{NMS2}(a)(c) and Fig.~\ref{NMS2}(b)(d), if one is interested to realize an appropriate metrology nearby the mechanical frequency, then one has to focus on the blue-sideband situation, in which the noise spectrum is nearly monotonically associated to the coupling strength. In the following section, it is shown that the coherent quantum noise cancellation should also stick to the blue-sideband condition.

\section{CQNC under balanced coupling}\label{CQNC}

\subsection{Cancellation of backaction noise}\label{balancedCQNC}

The quantum metrology based on CQNC technique targets on reducing the backaction noise~\cite{CQNC2,CQNC1}, where thermal noise and other technical noise sources can be avoidable in principle~\cite{CQNC3}, so as to realize an ultra-sensitive detection beyond the standard quantum limit. As for the noise spectrum in Eq.~(\ref{NoiseSpectrum}), SQL in Eq.~(\ref{SQL}) is a compromise of the backaction noise weighted with $G^2$ and the shot noise weighted with $1/G^2$. In this section, it is shown that in our double-mode optomechanical system, both the backaction noise and the background noise (independent on $G$) can be dramatically reduced by coupling the probe mode with an ancillary mode.

The core idea of the CQNC proposal lies in the coupling between the probe mode $b$ and the ancillary mode $c$. The extra noise induced by this coupling compensates the backaction noise on the mechanical oscillation with an opposite sign. The interaction between mode-$b$ and mode-$c$ is divided into the rotating-wave terms that can be realized by beam splitters (BS) and the counter-rotating terms that can be realized by optical parametric amplifiers (OPA). Mode-$c$ is tuned to be near-resonant with mode-$b$ to avoid the unnecessary interaction with mode-$a$. The Hamiltonian under the CQNC control can be generalized from the initial Hamiltonian in Eq.~(\ref{Hamiltonian}). In the rotating frame with respect to $H_d'=\omega_d(a^\dagger a+b^\dagger b+c^\dagger c)$, it reads
\begin{equation}\label{HCQNC}
\begin{aligned}
H'&=\Delta_aa^\dagger a+\Delta_b b^\dagger b+\Delta_c c^\dagger c+\frac{\omega_m}{2}(x^2+p^2)\\
&+iE(a^\dagger-a)+gx(a^\dagger b+ab^\dagger)+g_1(bc^\dagger+b^\dagger c)\\
&+g_2(bc+b^\dagger c^\dagger),
\end{aligned}
\end{equation}
where $c$ ($c^\dagger$) is the annihilation (creation) operator for the ancillary mode and $\Delta_c\equiv\omega_c-\omega_d\approx\Delta_b$ is the detuning between the ancillary cavity and the driving laser. The last two terms in Eq.~(\ref{HCQNC}) are used for coherent cancellation of noise, which can be tuned in experiments. The rotating-wave terms weighted with $g_1$ describe a passive BS mixing the two cavity modes; while the counter-rotating terms weighted with $g_2$ are denoted by an active down-conversion dynamics of the two modes through a nondegenerate OPA. The strength of the rotating-wave interaction caused by BS is $g_{1}=rc/L$, where $r$ is the reflectivity, $c$ is the speed of light, and $L$ is the cavity length. The strength of the counter-rotating interaction is $g_{2}=\Gamma lc/L$, where $l$ is the crystal length and $\Gamma$ is the gain parameter~\cite{CQNC2}.

Now the full set of the Heisenberg-Langevin equations describing the dynamics of the total system reads,
\begin{equation}\label{Langevin2}
	\begin{aligned}
		&\dot a=-i\Delta_aa-ig x b-\kappa_a a+\sqrt{2\kappa_a}a_{\rm in}+E,\\
		&\dot b=-i\Delta_bb-ig x a-ig_{1}c-ig_{2}c^\dagger-\kappa_bb+\sqrt{2\kappa_b}b_{\rm in},\\		&\dot c=-i\Delta_cc-ig_{1}b-ig_{2}b^\dagger-\kappa_cc+\sqrt{2\kappa_c}c_{\rm in},\\ 	&\dot  x=\omega_m p,\\
		&\dot p=-\omega_m x-g(a^\dagger b+ab^\dagger)-\gamma_m p+F_{\rm in},
	\end{aligned}
\end{equation}
where $c_{\rm in}$ is the vacuum noise operator for the ancillary mode satisfying $\langle c_{\rm in}(t)c^\dagger_{\rm in}(\tau)\rangle=\delta(t-\tau)$. Again, after the linearization procedure, the Heisenberg-Langevin equations for the fluctuation variables are
\begin{equation}\label{langevin3}
	\begin{aligned}
		&\dot a=-i\Delta_aa-\kappa_a a+\sqrt{2\kappa_a}a_{\rm in},\\
		&\dot b=-i\Delta_bb-iG x-i\frac{g_c}{2}c-i\frac{g_c}{2}c^\dagger-\kappa_bb+\sqrt{2\kappa_b}b_{\rm in},\\		&\dot c=-i\Delta_cc-i\frac{g_c}{2}b-i\frac{g_c}{2}b^\dagger-\kappa_cc+\sqrt{2\kappa_c}c_{\rm in},\\		&\dot x=\omega_m p,\\		&\dot p=-\omega_m x-G(b+b^\dagger)-\gamma_m p+F_{\rm in},
	\end{aligned}
\end{equation}
where the coupling strengthes are chosen as $g_{2}=g_{1}=g_c/2$ for a balanced configuration in this section (the imbalanced configuration will be further discussed in Sec.~\ref{unbalance}). This choice is irrespective to the fact that the driven mode-$a$ is again decoupled from all the interested modes. Similar to Eq.~(\ref{linearLanggevin}), it means that the measurement noise is immune to the fluctuations of the driven mode. The decoupling as well as the noise cancellation is ensured by the near-resonant condition $\Delta_c\approx\Delta_b$. From Eq.~(\ref{langevin3}), one can write the equations of motion for the field quadratures of the mechanical oscillator and the optical modes $b$ and $c$:
\begin{equation}\label{langevin7}
	\begin{aligned}
		&\dot x_b=\Delta_bp_b-\kappa_bx_b+\sqrt{2\kappa_b}x_{\rm in}^b, \\
		&\dot p_b=-\Delta_bx_b-\sqrt{2}G x-g_cx_c-\kappa_bp_b+\sqrt{2\kappa_b}p^b_{\rm in}, \\		&\dot x_c=\Delta_cp_c-\kappa_cx_c+\sqrt{2\kappa_c}x^c_{\rm in}, \\
		&\dot p_c=-\Delta_cx_c-g_cx_b-\kappa_cp_c+\sqrt{2\kappa_c}p^c_{\rm in},\\
		&\dot  x=\omega_mp, \\
		&\dot p=-\omega_m x-\sqrt{2}G x_b-\gamma_m p+F_{\rm in},
	\end{aligned}
\end{equation}
where $x_c\equiv(c+c^\dagger)/\sqrt{2}$, $p_c\equiv(c-c^\dagger)/\sqrt{2}i$, $x_{\rm in}^c\equiv(c_{\rm in}+c^\dagger_{\rm in})/\sqrt{2}$, and $p_{\rm in}^c\equiv(c_{\rm in}-c^\dagger_{\rm in})/\sqrt{2}i$.

\begin{figure}[htbp]
\centering
\includegraphics[width=0.9\linewidth]{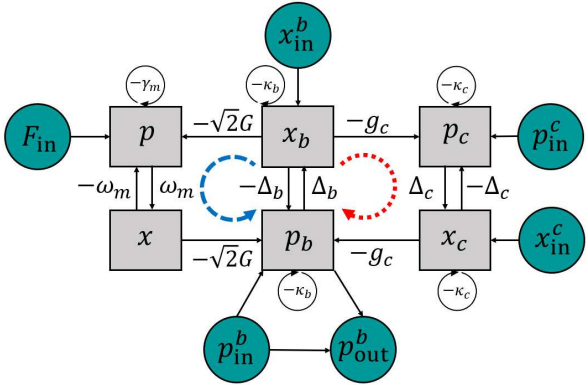}
\caption{(Color online) Flow chart for Eq.~(\ref{langevin7}) and the input-output theory in Eq.~(\ref{InOut}). Arrows point from the variables on the right side of Eq.~(\ref{langevin7}) to the relevant variables on its left side. The blue dashed curve describes the propagating path of the input noise $x_{\rm in}^b$ from the position quadrature $x_b$ through the variables $p$ and $x$ to the probed phase quadrature $p_b$. The red dotted curve denotes the assisted path by the ancillary mode having an opposite effect to compensate the blue one.} \label{FlowChart}
\end{figure}

A flow chart on Eq.~(\ref{langevin7}) is drafted in Fig.~\ref{FlowChart} to visualize how to build the antinoise coherent channel via the ancillary mode. In the linear-response regime, the output field is the sum of the individual contributions from the input signals and noises. In our protocol for the weak-force metrology, the to-be-measured external force acting on the mechanical oscillator formally generates an input field (including signal and noise) to the probe mode, which propagates to the output signal through the quadrature variables $p$, $x$, and $p_b$ in sequence. The backaction noise~\cite{CQNC1} $x_{\rm in}^b$ contributes to the output signal partially through the coupling between the probe mode $b$ and the mechanical mode and partially through the coupling between the probe mode $b$ and the ancillary mode $c$. The two paths are respectively distinguished by the blue dashed and the red dotted curves in Fig.~\ref{FlowChart}. The ancillary coherent channel described by the red curve plays a central role in CQNC. It is verified that under the blue-sideband condition $\Delta_c=-\omega_m$, the ancillary mode behaves effectively as a negative-frequency mechanical oscillator, equivalent to a mechanical oscillator with an effectively negative mass~\cite{CQNC1}. Intuitively, this condition shares a similar formation as to that for the single-mode CQNC strategy~\cite{CQNC1,CQNC2}. Note $\Delta_c$ in the single-mode case is the detuning between the ancillary mode and the unique system mode that is strictly resonant with the driving frequency. In our case, however, $\Delta_c$ is the detuning between the ancillary mode and the driving frequency. Our driven mode-$a$ is decoupled from the rest modes, then its frequency $\omega_a$ is more flexible in magnitude. Consequently the ancillary mode generates an extra backaction noise with an opposite sign to the initial one, which facilitates a destructive quantum interference. The output noise spectrum for the probe mode in Eq.~(\ref{NoiseSpectrum}) is therefore modified as
\begin{equation}\label{Sc}
\begin{aligned}
S^c(\omega)&=S_{\rm th}+\frac{g_c^2\kappa_c|\chi_c(\kappa_c-i\omega)|^2}{2G^2\Delta_c^2|\chi_m|^2}+
\frac{g_c^2\kappa_c|\chi_c|^2}{2G^2|\chi_m|^2}\\
&+\frac{1}{2}\Bigl |\frac{\sqrt{\kappa_b}\chi_b\Delta_b-\sqrt{\kappa_b}	\chi_b(g_c^2\chi_c+2G^2\chi_m)}{G\chi_m}\Bigr|^2\\		&+\frac{1}{2}\Big|\frac{1-2\kappa_b\chi_b+\Delta_b^2\chi_b^2-\Delta_b\chi_b^2(g_c^2\chi_c+2G^2\chi_m)}{2G\sqrt{\kappa_b}
\chi_b\chi_m}\Big|^2,
\end{aligned}
\end{equation}
where the susceptibilities of the probe field, the mechanical oscillator and the ancillary field are respectively defined as
\begin{subequations}\label{Susceptibilities}
	\begin{align}
		\chi_b\equiv &\frac{1}{\kappa_b-i\omega},\\
		\chi_m\equiv &\frac{\omega_m}{\omega_m^2-\omega^2-i\omega\gamma_m}, \label{chim} \\
		\chi_c\equiv &\frac{\Delta_c}{(\kappa_c-i\omega)^2+\Delta_c^2}. \label{chic}
	\end{align}
\end{subequations}
An extra backaction noise $g_c^2\chi_c$ emerges in the last two terms in Eq.~(\ref{Sc}), clearly showing its interference with the initial noise $2G^2\chi_m$. Then an ideal noise cancellation by destructive interference requires
\begin{equation}\label{MC}
	g_c^2\chi_c+2G^2\chi_m=0.
\end{equation}
For simplicity, the total coupling strength between the ancillary mode and the probe mode is set as
\begin{subequations}\label{MatchingCondition}
\begin{equation}\label{StrengthMatching}
g_c=\sqrt{2}G,
\end{equation}
which can be conveniently satisfied by modulating BS and OPA. Consequently it is demanded that $\chi_c=-\chi_m$. Then due to the definitions given in Eqs.~(\ref{chim}) and (\ref{chic}) and under the resolved sideband condition $\omega_m\gg\kappa_c$, the ancilla-mode frequency should satisfy
\begin{equation}\label{DetuningMatching}
	\Delta_c=-\omega_m,
\end{equation}
and meanwhile the linewidth of the ancillary mode is half of that of the mechanical oscillator
\begin{equation}\label{DecayMatching}
\kappa_c=\frac{\gamma_m}{2}.
\end{equation}
\end{subequations}

\begin{figure}[htbp]
\centering
\includegraphics[width=0.9\linewidth]{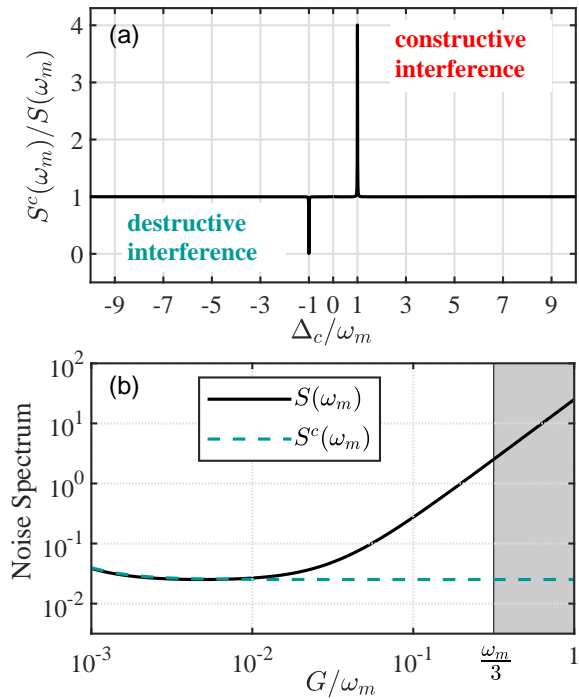}
\caption{(Color online) (a) Ratio of the measurement noise spectrum under the CQNC protocol and that with no control by varying the ancilla-mode detuning under $\Delta_b=\Delta_c$. (b) The measurement noise spectrums $S^c(\omega_m)$ and $S(\omega_m)$ as functions of coupling-strength $G$ in the presence and in the absence of CQNC under $\Delta_b=\Delta_c=-\omega_m$. The gray areas indicate the unstable regimes starting from $G=\omega_m/3$ determined by the Routh-Hurwitz criterion.  In both (a) and (b), the enhanced coupling strength between the probe mode and the mechanical oscillator is $G=0.2\omega_m$ and the relaxation rate of the ancillary mode is $\kappa_c=\gamma_m/2$ as given in Eq.~(\ref{DecayMatching}). The other parameters are set the same as those in Fig.~\ref{NMS2}. } \label{Ratio_and_CQNC}
\end{figure}

The effect from the quantum interference of the two coherent noisy channels (distinguished by the blue and red curves in Fig.~\ref{FlowChart}) on the output noise spectrum is shown in Fig.~\ref{Ratio_and_CQNC}(a) by the ratio $S^c(\omega_m)/S(\omega_m)$. Here we focus on the reduced noise spectrum around the mechanical-oscillator frequency $\omega_m$ by CQNC under the resonant condition $\Delta_b=\Delta_c$. The destructive interference at the blue-sideband $\Delta_c=-\omega_m$ and the constructive interference at the red-sideband $\Delta_c=\omega_m$ are respectively presented by the two extremely sharp patterns. This result justifies again the choice of the blue-detuning condition for the double-mode CQNC sensing.

Thus under the destructive-interference condition in Eq.~(\ref{MC}) for the backaction noise, the full noise spectrum becomes
\begin{eqnarray}\nonumber
&& S^{c}(\omega)=S_{\rm th}+\kappa_c\frac{|\chi_c(\kappa_c-i\omega)|^2+\Delta_c^2|\chi_m|^2}{\Delta_c^2|\chi_m|^2}\\		&+&\frac{\bigl[(\Delta_b^2-\omega^2+\kappa_b^2)^2+4\kappa_b^2\omega^2\bigr](\omega^2\gamma_m^2+\Delta^2)}
{8G^2\kappa_b\omega^2_m(\omega^2+\kappa_b^2)}. \label{ScWithoutBackaction}
\end{eqnarray}
In comparison to the noise spectrum without control in Eq.~(\ref{NoiseSpectrum}), the thermal and the shot noises remain invariant, the backaction noise for the mechanical oscillator is cancelled, and the extra background noise induced by mode-$c$ turns out to replace the previous one proportional to the detuning $\Delta$ [the second term in Eq.~(\ref{NoiseSpectrum})].

In Fig.~\ref{Ratio_and_CQNC}(b), the numerical simulations of $S(\omega_m)$ (see the black solid line) and $S^c(\omega_m)$ (see the green dashed line) for CQNC are shown at the working point $\Delta_b=\Delta_c=-\omega_m$ by varying the coupling strength $G$ between the probe and the mechanical modes. When increasing the coupling strength $G$, $S(\omega_m)$ will firstly decrease and then rebound. The lower bound of $S(\omega_m)$, which is located at the optimized point $G=G_L$ determined by Eq.~(\ref{SQLG}), is the standard quantum limit $S_L(\omega_m)$ given by Eq.~(\ref{SQL}). In the weak-coupling regime, the spectrum under the CQNC strategy $S^c(\omega_m)$ is almost the same as $S(\omega_m)$. The control based on the destructive interference over the backaction noise begins to take effect for a moderate coupling strength ($\sim 10^{-2}\omega_m$). When $G$ is over the SQL point, $S^c(\omega_m)$ becomes significantly lower than $S(\omega_m)$ and will never rebound. That means our CQNC protocol can effectively reduce the measurement noise of the backaction part, especially under strong coupling regime. Note in our system, the thermal noise acting on the mechanical oscillator in the limit of sufficiently large driving powers is negligible. And in both free and control situations (Fig.~\ref{NMS2} and Fig.~\ref{Ratio_and_CQNC}), the average population for the thermal noise is fixed to distinguish the functionality of CQNC on reducing the backaction noise.

Furthermore, the consistency of lower bounds for both $S^c(\omega_m)$ and $S(\omega_m)$ shown in Fig.~\ref{Ratio_and_CQNC}(b) could be illustrated analytically. Substituting Eqs.~(\ref{DetuningMatching}) and (\ref{DecayMatching}) into Eq.~(\ref{ScWithoutBackaction}) and omitting the shot noise (scaling as $G^{-2}$ and then being negligible for a large coupling strength), one can find that $S_L^c(\omega)\approx S_{\rm th}+\kappa_c(\omega_m^2+\omega^2+\kappa_c^2)/\omega_m^2$. Regarding the resonant condition $\omega=\omega_m$ and noting $\omega_m\gg\gamma_m, \kappa_c$, it converges to the SQL solution in Eq.~(\ref{SQL}): $S_L^c(\omega_m)\approx S_L(\omega_m)\approx S_{\rm th}+2\kappa_c$.

\subsection{Stability Analysis}

Generally, two main aspects need to be considered about the stability of an optomechanical system: a stable linear response ensured by the Routh-Hurwitz criterion, and a stable optical spring ensured by a positive effective mechanical-frequency and a positive effective-damping-rate. First, the Routh-Hurwitz criterion has already been considered in appendix~\ref{StabilityCondition}, where the stability is translated to a constraint for the coupling strength. The gray area in Fig.~\ref{Ratio_and_CQNC}(b) indicates an unstable regime beyond a coupling-strength threshold. In the stable regime, still there is a reduction about two orders of magnitude for the backaction noise.

The response of the mechanical position to the radiation pressure characterizes a second stability criterion by introducing a mechanical frequency shift $\delta\omega_m$ (named rigidity in~\cite{Rigidity1,Rigidity2}) and an optomechanical damping rate $\gamma_{\rm opt}$. According to Sec.~V.B in the review of Ref.~\cite{cavityOpto}, both the shifted mechanical frequency $\omega_{\rm eff}^2\equiv\omega_m^2+2\omega\delta\omega_m$ and the effective mechanical damping $\gamma_{\rm eff}\equiv\gamma_m+\gamma_{\rm opt}$ should be positive to ensure the stability of system. It is found that the blue-detuning driving as considered in our protocol may introduce a negative damping, that could lead to antidamping or even instability~\cite{cavityOpto}. The possibility of instability can be avoided via certain strategies, such as employing two frequency-offset laser fields~\cite{DoubleOpticalSpring} or utilizing birefringence~\cite{birefringence}, etc. Our CQNC protocol, however, could offset the backaction noise and simultaneously promote a stable optical spring without any extra control.

Regarding the double-mode system without the ancillary mode $c$, the modified susceptibility in Eq.~(\ref{susceptibility}) could be reexpressed by the decoupled susceptibility and an optical rigidity $\Sigma(\omega)$ due to the coupling between the system modes and the mechanical oscillator:
\begin{subequations}
\begin{equation}\label{ModifiedSusceptibility}
    \chi(\omega)=\left[\frac{\omega_m^2-i\omega\gamma_m-\omega^2}{\omega_m}+\Sigma(\omega)\right]^{-1},
\end{equation}
where
\begin{equation}\label{modification}
    \Sigma(\omega)=-\frac{2G^2\Delta_b}{(\kappa_b-i\omega)^2+\Delta_b^2}.
\end{equation}
\end{subequations}
Then due to the general definition,
\begin{equation}\label{Sigmadefinition}
    \Sigma(\omega)\equiv\frac{1}{\omega_m}\left(2\omega\delta\omega_m-i\omega\gamma_{\rm opt}\right),
\end{equation}
we have
\begin{equation}
\begin{aligned}
\delta\omega_m&=-\frac{G^2\Delta_b(\Delta_b^2+\kappa_b^2-\omega^2)\omega_m}{\left[\Delta_b^4+2\Delta_b^2(\kappa_b^2-\omega^2)
+(\kappa_b^2+\omega^2)^2\right]\omega}, \\
\gamma_{\rm opt}&=\frac{4G^2\Delta_b\kappa_b\omega_m}{4\kappa^2\omega^2+(\Delta_b^2+\kappa_b^2-\omega^2)^2}.
\end{aligned}
\end{equation}
Under the near-resonant condition, the magnitude of the frequency shift is much lower than the original mechanical frequency $\left|\delta\omega_m\right|\ll\omega_m$, which ensures a positive effective mechanical frequency $\omega_{\rm eff}$. Under the conditions of blue-sideband $\Delta_b=-\omega_m$ and near-resonance $\omega\approx\omega_m$, however, it is found that
\begin{equation}\label{OriginalGamma}
\gamma_{\rm eff}=\gamma_m+\frac{4G^2\Delta_b\kappa_b\omega_m}{4\kappa_b^2\omega^2+(\Delta_b^2+\kappa_b^2-\omega^2)^2}
\approx\gamma_m-\frac{G^2}{\kappa_b}<0
\end{equation}
unless in the weak-coupling regime that $G^2<\kappa_b\gamma_m$. In another word, the strong coupling between the probe mode and the mechanical mode required for noise cancellation will inevitably renders a negative damping-rate since it  continuously heats up the mechanical oscillator.

This undesirable effect can be avoided by coupling the probe mode to the ancillary mode. The CQNC protocol can provide a cooling mechanism for the optomechanical interaction. A standard treatment following the Heisenberg-Langevin equation~(\ref{langevin7}) that describes the dynamics of the system under control yields a new optical rigidity
\begin{equation}\label{susceptibility_control}
\Sigma^c(\omega)=-\frac{2G^2\Delta_b}{(\kappa_b-i\omega)^2+\Delta_b^2-g_c^2\Delta_b\chi_c}.
\end{equation}
The exact expression for the modified extra damping rate is rather tedious and lack of a physical insight. Consider $\kappa_c,\gamma_m\ll G,g_c,\omega_m,\omega$, one can find that
\begin{equation}
\gamma_{\rm opt}^{c}\approx -2\kappa_c+\frac{4\kappa_b\kappa_c^2}{G^2}
\end{equation}
upper to the third-order of $\kappa_c$ and $\gamma_m$. Recalling the anti-noise condition in Eq.~(\ref{DecayMatching}) to completely compensate the radiation-pressure-caused noise, $\gamma_m=2\kappa_c$, we find that the effective mechanical damping rate
\begin{equation}\label{Gamma_balance}
\gamma_{\rm eff}^c\equiv\gamma_m+\gamma_{\rm opt}^c\approx\frac{4\kappa_b\kappa_c^2}{G^2}>0.
\end{equation}
In comparison to the result in Eq.~(\ref{OriginalGamma}), the coupling between the probe mode and the ancillary mode therefore pull the system back into the stable regime in the presence of the blue-detuning driving. The extra coupling establishes essentially a phonon-leaking channel for the mechanical oscillation by the energy-exchanging interaction [see the rotating-wave terms in Eq.~(\ref{HCQNC})]. The motion of the mechanical oscillator then tends to be stable rather than vibrating violently. Hereafter, we break the balance between the rotating-wave terms $b^\dagger c+bc^\dagger$ (in charge of cooling) and the counter-rotating terms $b^\dagger c^\dagger+bc$ (in charge of heating) to dig out more physics associated to them.

\section{CQNC under imbalanced coupling}\label{unbalance}

With an imbalanced coupling between the probe mode and the ancillary mode, the Heisenberg-Langevin equation~(\ref{langevin7}) that describes the CQNC dynamics of the system is modified to
\begin{equation}\label{langevin_g1_g2}
	\begin{aligned}
		&\dot x_b=\Delta_bp_b+g_np_c-\kappa_bx_b+\sqrt{2\kappa_b}x_{\rm in}^b, \\
		&\dot p_b=-\Delta_bx_b-\sqrt{2}G x-g_cx_c-\kappa_bp_b+\sqrt{2\kappa_b}p^b_{\rm in}, \\		&\dot x_c=\Delta_cp_c+g_cp_b-\kappa_cx_c+\sqrt{2\kappa_c}x^c_{\rm in}, \\
		&\dot p_c=-\Delta_cx_c-g_cx_b-\kappa_cp_c+\sqrt{2\kappa_c}p^c_{\rm in},\\
		&\dot  x=\omega_mp, \\
		&\dot p=-\omega_m x-\sqrt{2}G x_b-\gamma_m p+F_{\rm in},
	\end{aligned}
\end{equation}
where $g_c\equiv g_1+g_2$ and $g_n\equiv g_1-g_2$. In this subsection, we fix the full strength of the anti-noise channel, i.e. $g_c=g_1+g_2=\sqrt{2}G$ [see Eq.~(\ref{MC})] and change the relative strength $g_n$. The rigidity expressed by Eq.~(\ref{susceptibility_control}) then turns out to be:
\begin{equation}\label{Sigmac}
	\begin{aligned}
&\Sigma^c(\omega)=\Bigl[2G^2\Delta_c\chi_b^2(\Delta_b+ig_n^2\chi_{cp}^{-1}\chi_{cn}\chi_c)\Bigr]\\
&\times\Bigl\{-\Delta_c(1+\Delta_b^2\chi_b^2+g_ng_c\chi_b\chi_{cn})+\chi_b\chi_c\bigl[g_c^2\Delta_b\Delta_c\chi_b\\
&+ig_n^2(g_c^2-\Delta_b\Delta_c)\chi_{cp}^{-1}\chi_{cn}\chi_b+g_ng_c(\Delta_c^2\chi_{cn}-\chi_{cn}^{-1})\bigr]\Bigr\}^{-1},
	\end{aligned}
\end{equation}
where $\chi_{cn}\equiv1/(\kappa_c-i\omega)$ and $\chi_{cp}=1/(\kappa_c+i\omega)$. It can be used to obtain the $g_n$-dependent effective damping rate for the mechanical mode $\gamma_{\rm eff}^c$ and CQNC-sensing spectrum $S^c(g_n,\omega_m)$ via the above standard procedure.

\begin{figure}[htbp]
	\centering
	\includegraphics[width=0.5\linewidth]{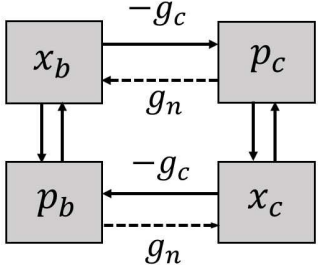}
	\caption{The flow chart between the probe mode and the ancillary mode under imbalanced coupling. The solid line indicates the original channel to compensate the original backaction noise (see the right side of Fig.~\ref{FlowChart}) and the dash lines indicate the back-flows induced by $g_n$.}\label{anti_noise_path}
\end{figure}

\begin{figure}[htbp]
	\centering
    \includegraphics[width=0.9\linewidth]{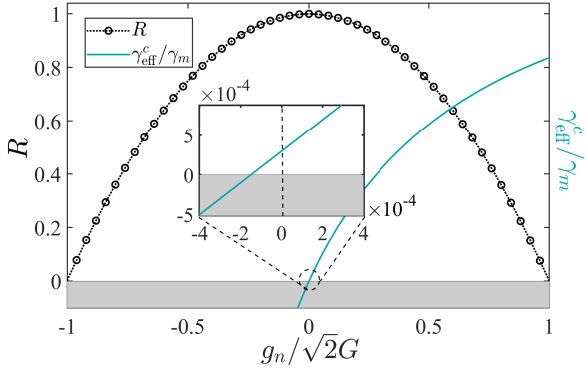}
	\caption{The ratios of both noise cancellation $R$ and $\gamma_{\rm eff}^c/\gamma_m$ as functions of the coupling $g_n$ normalized by $\sqrt{2}G$ with a fixed $g_c=\sqrt{2}G$. The inset shows $\gamma_{\rm eff}^c/\gamma_m$ around the balanced point ($g_n=0$). The gray areas indicate the unstable regimes. The parameters are set as $G=0.2\omega_m$, $\Delta_b=\Delta_c=-\omega_m$, $\kappa_b=10^{-2}\omega_m$, $\gamma_m=1.2\times10^{-3}\omega_m$ and $n_{\rm th}=10$.}\label{ConcelRate_and_Gamma}
\end{figure}

As shown in the flow chart of Fig.~\ref{anti_noise_path}, the nonzero $g_n$ would bring about a back-flow along the assisted path having the opposite effect to compensate the original backaction noise. It is expected to weaken the noise cancellation effect by CQNC, which could be estimated by the following ratio:
\begin{equation}\label{cancel_rate}
R\equiv1-\frac{S^c(g_n,\omega_m)}{S(\omega_m)}.
\end{equation}
Accordingly, the CQNC strategy works when $R>0$. The ratios of both noise cancellation $R$ and $\gamma_{\rm eff}^c/\gamma_m$ are plotted in Fig.~\ref{ConcelRate_and_Gamma} as functions of the normalized $g_n$. With a fixed $g_c$ and in the absence of the counter-rotating interaction $g_2$, i.e., the right limit of Fig.~\ref{ConcelRate_and_Gamma} with $g_n=g_1=g_c=\sqrt{2}G$, it is then found that $\gamma_{\rm opt}^c\approx-\kappa_c/3$ and then the full damping rate $\gamma_{\rm eff}^c\approx5\kappa_c/3$ is much larger than that in the balanced-coupling case. It distinguishes the role played by the rotating-wave interaction with the ancillary mode as a quantum refrigerator, continuously extracting the energy from the pushed mechanical mode by a faster decay rate and leaving the full system in a more stable state. Yet in the same time, it will completely cancel the antinoise path in Fig.~\ref{anti_noise_path} to destroy the destructive intercurrence demanded by CQNC, i.e., $R(g_n=g_c)=0$. In contrast, a dominant counter-rotating interaction $g_2$, i.e., $g_n<0$, yields a negative $\gamma_{\rm eff}^c$ but a positive $R$, where the counter-rotating coupling heats up the mechanical mode and pushes the full system into the unstable regime. When $g_n\geq0$, i.e., the rotating-wave terms dominate while the counter-rotating terms simultaneously present, an increasing $g_n$ acts as an opposite role in building the anti-noise channel yielding a decreasing $R$ and a positive $\gamma_{\rm eff}^c$ to stabilize the full system. The distinct effects for the rotating-wave and counter-rotating wave terms are consistent with those about the cooling condition of an open-quantum-system in a finite-temperature environment by measurements~\cite{ZhangandJing}. Note that the balanced coupling $g_n=0$ described by the inset in Fig.~\ref{ConcelRate_and_Gamma} is stable, which is consistent with Eq.~(\ref{Gamma_balance}), and also maximizes $R$.

\begin{figure}[htbp]
\centering
\includegraphics[width=0.9\linewidth]{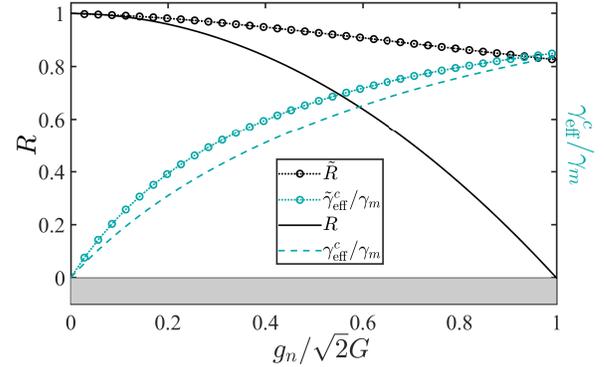}
\caption{The ratios of both noise cancellation $R$ and $\gamma_{\rm eff}^c/\gamma_m$ as functions of the coupling $g_n$ normalized by $\sqrt{2}G$ with a fixed $g_c=\sqrt{2}G$ (the lines with no circles) and with a fixed $g_2=\sqrt{2}G$ (the lines with circles). The other parameters are set the same as those in Fig.~\ref{ConcelRate_and_Gamma}. }\label{CancelR_and_Gamma_fixed_g2}
\end{figure}

The complementary results of the noise cancellation ratio and the effective damping rate in Fig.~\ref{ConcelRate_and_Gamma} can be partially relieved by enhancing the full coupling strength of the antinoise path, which is actually determined by $g_c-g_n$ due to Fig~\ref{anti_noise_path}. In Fig.~\ref{CancelR_and_Gamma_fixed_g2}, it is found by fixing the counter-rotating coupling strength $g_2=(g_c-g_n)/2=\sqrt{2}G/2$ and enhancing $g_n$ as well as the rotating-wave coupling $g_1$, we can have a more robust antinoise channel and a more stable sensing system (see the two line with circles, where the enhanced quantities are marked by tildes) in the same time. Comparing to the previous results (see the two line with no circles), now the amplified rotating-wave coupling dramatically reduces the decline of the noise-cancellation ratio with an increasing $g_n$ while holding almost the same stability of the full system. When $g_n$ approaches $\sqrt{2}G$, our CQNC protocol can still offset over $80\%$ of the original backaction noise in comparison to the balanced situation analysed in Sec.~\ref{balancedCQNC}.

\begin{table}[htbp]
\caption{The noise-cancellation ratio $\tilde{R}(\omega=\omega_m)$, the effective damping rate $\tilde{\gamma}_{\rm eff}^c(\omega=\omega_m)/\gamma_m$, the stable frequency-range for the measurement around the mechanical frequency $[\omega_m-\Delta\omega,\omega_m+\Delta\omega]$ with positive $\tilde{\gamma}_{\rm eff}^c$ and the coupling-strength threshold $G_{\rm max}$ based on the Routh-Hurwitz criterion versus the normalized $g_n$ with a fixed $g_2=\sqrt{2}G$. }
\begin{ruledtabular}
\begin{tabular}{cccccccc}
 $g_n/\sqrt{2}G$ & $\Delta\omega/\omega_m$ & $\tilde{R}$ & $\tilde{\gamma}_{\rm eff}^c/\gamma_m$ & $G_{\rm max}/\omega_m$\\
\hline
$0.1$ & 0.02 & 0.31 & 0.99 & 1.5\\
$0.2$ & 0.04 & 0.49 & 0.98 & 1.1\\
$0.3$ & 0.06 & 0.62 & 0.97 & 0.85\\
$0.4$ & 0.08 & 0.70 & 0.95 & 0.72\\
$0.5$ & 0.10 & 0.76 & 0.93 & 0.63\\
\end{tabular}
\end{ruledtabular}
\label{FrequencyRange}
\end{table}

Up to now, we focus on the resonant or near-resonant situation. In Table~\ref{FrequencyRange}, one can find the working range for the measurement frequencies under the enhanced protocol observed in Fig.~\ref{CancelR_and_Gamma_fixed_g2}, where the rotating-wave coupling strength between the probe mode and the ancillary mode increases with a fixed counter-rotating coupling strength. It is interesting to see the width of the working frequency range for the weak-forcing sensing is almost proportional to $g_n$, the imbalance between $g_1$ and $g_2$. And the resonant noise cancellation ratio and the effective damping rate still hold a complementary relation and the upper-bound of the coupling strength decreases with the increasing $g_n$ as well as the full coupling strength. 

\section{Physical Realization}\label{PhyRealize}

\begin{figure}[htbp]
\centering
\includegraphics[width=0.9\linewidth]{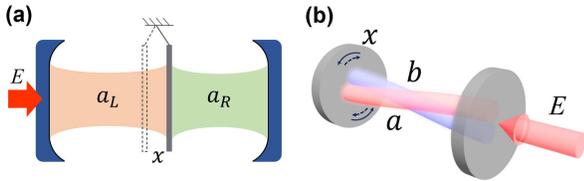}
\caption{(Color online). (a) Diagram of the mirror-in-the-middle optomechanical system comprising a Fabry-P\'erot cavity with a high-reflectivity mirror mounted in the middle. The two optical modes $a_L$ and $a_R$ are coupled by the middle-mirror displacement. (b) Diagram of the torsional optomechanical system, in which an uniaxial medium is inserted. The cavity is weakly twisted with the torsional oscillation of the end mirror, which causes the mixing of the ordinary mode $a$, the extraordinary mode $b$ and the mechanical mode $x$. The ordinary-mode frequency is larger than the extraordinary-mode frequency.} \label{model}
\end{figure}

We now consider accessible implementations of our protocol. The initial Hamiltonian in Eq.~(\ref{Hamiltonian}) could be realized by two types of optomechanical systems with two optical modes.

The Hamiltonian of a prototypical system comprising a Fabry-P\'erot cavity with a mirror of high-reflectivity mounted in the middle~\cite{three-mode,squeezing1} [see Fig.~\ref{model}(a)] reads,
\begin{eqnarray}\nonumber
H&=&(\omega+fx)a^\dagger_La_L+(\omega-fx)a^\dagger_Ra_R\\ &-& J(a^\dagger_La_R+a^\dagger_Ra_L),  \label{MiddleMirrorHamiltonian1}
\end{eqnarray}
where $\omega$ is the frequency of the two subcavities, $f$ is the frequency shift per unit length and $a_L$ ($a_R$) is the annihilation operator for the left (right) cavity mode and $J$ indicates the coupling strength between the left and the right cavity modes. Note $f\approx-\omega/L$ in the high-reflectivity limit, where $L$ is the cavity length.

This Hamiltonian could be diagonalized by two normalized modes through the unitary transformation $a=(a_L+a_R)/\sqrt{2}$ and $b=(a_L-a_R)/\sqrt{2}$. The normalized modes will have distinct frequencies due to the mutual coupling between cavity modes. Taking account the free energy of vibrating mirror and the external driving over mode-$a$ into consideration, one can modify the Hamiltonian in Eq.~(\ref{MiddleMirrorHamiltonian1}) into
\begin{eqnarray}\nonumber
H'&=&\omega_aa^\dagger a+\omega_bb^\dagger b+\frac{\omega_m}{2}(x^2+p^2)\\ \label{MiddleMirrorEigenfrequency2}
&+& g_mx(a^\dagger b+ab^\dagger)+iE(a^\dagger e^{-i\omega_dt}-ae^{i\omega_dt}),
\end{eqnarray}
where $\omega_{a,b}\equiv\omega\pm\sqrt{f^2x^2+J^2}$, $g_m=f$ and $|E|=\sqrt{P_{\rm in}\kappa_a/\omega_d}$ is the driving strength.

Another experimental platform is a weakly twisted optomechanical cavity~\cite{twisting} with a birefringent medium inside [see Fig.~\ref{model}(b)]. The two optical modes in this fire-new setup is the ordinary mode (marked as $a$) with a higher frequency $\omega_a$ and the extraordinary mode (marked as $b$) with a lower frequency $\omega_b$. The torsional oscillation of the back mirror renders the mechanical mode coupled to the optical modes via permittivity tensor modulation. With the ordinary mode being driven by an external driving laser and the extraordinary mode being probed, the Hamiltonian of the system can be written as
\begin{eqnarray}\nonumber
H&=&\omega_aa^\dagger a+\omega_bb^\dagger b+\frac{\omega_m}{2}(x^2+p^2)+ g_t x(a^\dagger b+ab^\dagger)\\ \label{twistedHamiltonian}
&+&iE(a^\dagger e^{-i\omega_d t}-ae^{i\omega_d t}).
\end{eqnarray}
Here the coupling strength of the twisted optomechanical cavity is expressed by
\begin{equation}\label{g}
g_t=-\frac{c}{16L\sqrt{n_en_o}}\left(\frac{1}{n_o^2}-\frac{1}{n_e^2}\right),
\end{equation}
where $c$ is the light speed, $L$ is the cavity length, and $n_o$ and $n_e$ are respectively the refractive indices of the ordinary and the extraordinary optical modes.

The noise spectral function for the weak-force measurement in SI unit of~$\rm N^2/Hz$ could be obtained by rescaling the noise spectrum via $\tilde{\mathcal{S}}(\omega)\equiv\hbar m\omega_m\mathcal{S}(\omega)$~\cite{CQNC2}, where $\mathcal{S}(\omega)$ could be $S(\omega)$, $S^c(\omega)$, $S_L(\omega)$ or $S^c_L(\omega)$ in certain conditions. In a typical optomechanical system, the mechanical frequency $\omega_m=2\pi\times10.56$MHz, the mechanical-oscillator mass $m=48$pg and the decay rate of the probe-mode $\kappa_b=2\pi\times200$kHz~\cite{cooling2}. Regarding the linewidth of the ancillary cavity and the condition in Eq.~(\ref{DecayMatching}), a comparatively large damping rate (taken as $\gamma_m=1.2\times10^{-3}\omega_m$ here) of the mechanical mode and a high-$Q$ ancillary cavity are on demand. The former could be conveniently realized by setting the mechanical oscillator in ambient environments rather than in a low vacuum~\cite{TorqueSensor}. The latter is accessible in the low-loss ($\kappa_c/2\pi\sim100$Hz) microwave cavities~\cite{AmplificationAndInstability,HighQCavity1,HighQCavity2}. Compared with the result with no control $\tilde{S}(\omega_m)=1.5\times10^{-28}{\rm N^2/Hz}$, the sensitivity under CQNC $\tilde{S}^c(\omega_m)=3.5\times10^{-30}{\rm N^2/Hz}$ is enhanced about two orders in magnitude at the resonant situation.

A twisted optomechanical system having cantilever nanomechanical resonators directly integrated within an optical nanocavity is proposed to detect a torque with a sensitivity of $1.2\times 10^{-20}~{\rm Nm/\sqrt{Hz}}$~\cite{TorqueSensor}. With the typical parameters (the mechanical frequency $\omega_m=2\pi\times4.9$MHz, the oscillator mass $m=427$fg, the mechanical quality factor $Q_m=\omega_m/\gamma_m=21$ and the mirror length $r=7.5~{\mu m}$) used in Ref.~\cite{TorqueSensor}, the improved sensitivity for weak-torque $\tilde{S}_{tor}(\omega_m)\equiv r\sqrt{\hbar m\omega_mS_L^c(\omega_m)}$ is found to be $7.75\times10^{-21}\rm Nm/\sqrt{Hz}$ under our CQNC strategy, also showing a better performance in torque detection.

\section{Conclusion}\label{conclusion}

In this work, we have proposed a coherent-quantum-noise-cancellation based metrology scheme in an optomechanical system consisting of two optical modes with distinct frequencies and a mechanical mode. It is found that to build up the antinoise channel for optimizing the weak-force sensitivity on the mechanical mode, the conventional CQNC metrology can be generalized by driving the high-frequency optical mode, probing the low-frequency optical mode and coupling the probe mode to a near-resonant ancillary mode. Under this deliberated and asymmetrical configuration, the standard quantum limit can be broken through by eliminating the initial backaction noise as well as remarkably reducing the entire noise-level. Moreover, we have investigated the details of interaction between the probe mode and the ancillary mode. It is important to find that a proper ratio between the rotating-wave coupling term and the counter-rotating term can be used to simultaneously realize the backaction-noise suppression through destructive interference and stabilize the full linear system.

Our proposal is accessible in both theory and experiment. The nanomechanical system is analyzed in the standard linearization process by following the linear-response theorem. In the Heisenberg-Langevin equation and the effective Hamiltonian, the strong coupling between the optical modes and the mechanical mode is crucial to realize the high-precision metrology and also constrained by the Routh-Hurwitz criterion and the stable optical spring requirement. Two experimental platforms for the nanomechanical systems, the membrane-in-the-middle setup and the twisted-cavity-based weak-torque detector, are found to be physical realizations for our metrology protocol. With a proper coupling strength, we have obtained dramatic improvements in terms of the rescaled sensitivity on the weak force or the weak torque.

\section*{Acknowledgments}

We acknowledge grant support from the National Science Foundation of China (Grants No. 11974311 and No. U1801661), Zhejiang Provincial Natural Science Foundation of China under Grant No. LD18A040001, and the Fundamental Research Funds for the Central Universities (No. 2018QNA3004).

\appendix

\section{The constraint on driving power by the linear-system stability}\label{StabilityCondition}

This appendix is about the Routh-Hurwitz criterion for constraining the pumping/driving power to ensure the linear-system stability. We first consider the linearized Heisenberg-Langevin equation~(\ref{linearLanggevin}) describing the free evolution of the system. The solution stability of the relevant ordinary differential equations can be determined by the eigenvalues of the associated Jacobian matrix. The solutions are stable if and only if the real part of every eigenvalue is negative. The differential equations in Eq.~(\ref{linearLanggevin}) decoupled from the driven mode are
\begin{equation}\label{langevin6}
	\begin{aligned}
		&\dot b=-i\Delta_bb-ig\alpha x-\kappa_b b+\sqrt{2\kappa_b}b_{\rm in}, \\
		&\dot b^\dagger=i\Delta_bb^\dagger+ig\alpha^* x-\kappa_b b^\dagger+\sqrt{2\kappa_b}b_{\rm in}^\dagger, \\
		&\dot  x=\omega_m p, \\
		&\dot p=-\omega_m x-g(\alpha b^\dagger+\alpha^*b)-\gamma_m p+F_{\rm in}.
	\end{aligned}
\end{equation}
It can be expressed in a compact form:
\begin{equation}\label{W}
	\dot W=A\cdot W+B,
\end{equation}
where $W$ and $B$ are column vectors
\begin{equation}\label{WAB}
	W=[b,b^\dagger, x,p]^{T},\quad B=[\sqrt{2\kappa_b}b_{\rm in}, \sqrt{2\kappa_b}b_{\rm in}^\dagger,0,F_{\rm in}]^T,
\end{equation}
and $A$ is a Jacobian matrix
\begin{equation}\label{A}
	A=\begin{pmatrix}-i\Delta_b-\kappa_b&0&-ig\alpha&0\\
	0&i\Delta_b-\kappa_b&ig\alpha^*&0\\
	0&0&0&\omega_m\\
	-g\alpha^*&-g\alpha&-\omega_m&-\gamma_m\\
	\end{pmatrix}.
\end{equation}
The characteristic function of $A$ satisfies $a_0\lambda^4+a_1\lambda^3+a_2\lambda^2+a_3\lambda+a_4=0$, where
\begin{equation}\label{character}
	\begin{aligned}
		&a_0=1,\\
		&a_1=\gamma_m+2\kappa_b,\\
		&a_2=\Delta_b^2+2\gamma_m\kappa_b+\kappa_b^2+\omega_m^2,\\
		&a_3=\gamma_m(\Delta_b^2+\kappa_b^2)+2\kappa_b\omega^2_m,\\
		&a_4=-2g^2|\alpha|^2\Delta_b\omega_m+\omega^2_m(\Delta_b^2+\kappa_b^2).
	\end{aligned}
\end{equation}
The Routh-Hurwitz stability criterion would constrain the system parameters through the Hurwitz determinants
\begin{equation}\label{determinant}
	\begin{aligned}
		D_1&=a_1, \quad
		D_2=\begin{vmatrix}a_1&a_3\\a_0&a_2\end{vmatrix}, \\
		D_3&=\begin{vmatrix}a_1&a_3&0\\a_0&a_2&a_4\\0&a_1&a_3\end{vmatrix}, \quad	D_4=\begin{vmatrix}a_1&a_3&0&0\\a_0&a_2&a_4&0\\0&a_1&a_3&0\\0&a_0&a_2&a_4\end{vmatrix}.
	\end{aligned}
\end{equation}
Note $D_1$ is positive. Thus according to the Routh-Hurwitz criterion, if and only if the following sequence of the determinants of its principal submatrixs, i.e. $D_i (i=2,3,4)$, are all positive, then the real parts of the eigenvalues of $A$ are all negative. A straightforward calculation shows that for $\Delta_b>0$  (the red-detuning case),
\begin{equation}\label{a2_1}
	|\alpha|^2<\frac{(\Delta_b^2+\kappa_b^2)\omega_m}{2|\Delta_b|g^2},
\end{equation}
and for $\Delta_b<0$ (the blue-detuning case)
\begin{equation}\label{a2_2}
	|\alpha|^2>-\frac{(\Delta_b^2+\kappa_b^2)\omega_m}{2|\Delta_b|g^2}.
\end{equation}

Next we include the extra interaction between the probe mode and the ancilla mode, which will modify the previous constrain condition for the driving power measured by $|\alpha|$. The relevant differential equations due to Eq.~(\ref{langevin3}) can be expressed by
\begin{equation}\label{Wc}
\dot W_c=A_c\cdot W_c+B_c,
\end{equation}
where
\begin{subequations}
	\begin{align}
		W_c=&[b,b^\dagger,c,c^\dagger,x,p]^{T}, \\
		B_c=&[\sqrt{2\kappa_b}b_{\rm in},\sqrt{2\kappa_b}b_{\rm in}^\dagger,\sqrt{2\kappa_c}c_{\rm in},\sqrt{2\kappa_c}c_{\rm in}^\dagger,0,F_{\rm in}]^T, \\		A_c=&
		\begin{pmatrix}\begin{smallmatrix}-i\Delta_b-\kappa_b&0&-ig_{1}&-ig_{2}&-iG&0\\
			0&i\Delta_b-\kappa_b&ig_{2}&ig_{1}&iG&0\\
			-ig_{1}&-ig_{2}&-i\Delta_c-\kappa_c&0&0&0\\
			ig_{2}&ig_{1}&0&i\Delta_c-\kappa_c&0&0\\
			0&0&0&0&0&\omega_m\\
			-G&-G&0&0&-\omega_m&-\gamma_m\\
		\end{smallmatrix}\end{pmatrix}.
	\end{align}
\end{subequations}
Then under the balanced condition that $g_{1}=g_{2}=g_c/2$, the characteristic function for the Jacobian matrix $A_c$ satisfies $a_0\lambda^6+a_1\lambda^5+a_2\lambda^4+a_3\lambda^3+a_4\lambda^2+a_5\lambda+a_6=0$, where the coefficients are
\begin{widetext}
\begin{equation}\label{character}
\begin{aligned}
	a_0=&1, \\
	a_1=&\gamma_m+2(\kappa_b+\kappa_c), \\	a_2=&\omega_m^2+\Delta_b^2+\Delta_c^2+2(\gamma_m\kappa_c+\gamma_m\kappa_b+\kappa_c\kappa_b)+(\kappa_b+\kappa_c)^2, \\	a_3=&\Delta_b^2(\gamma_m+2\kappa_c)+\Delta_c^2(\gamma_m+2\kappa_b)+2\omega_m^2(\kappa_b+\kappa_c)
+\gamma_m(\kappa_b+\kappa_c)^2+2\kappa_b\kappa_c(\gamma_m+\kappa_b+\kappa_c), \\		a_4=&-\Delta_b(2G^2\omega_m+g_c^2\Delta_c)+\Delta_c^2\omega_m^2+\Delta_b^2\omega_m^2+\Delta_b^2\Delta_c^2
+\omega_m^2[(\kappa_b+\kappa_c)^2+2\kappa_b\kappa_c]+\Delta_b^2\kappa_c(2\gamma_m+\kappa_c) \\		&+\Delta_c^2\kappa_b(2\gamma_m+\kappa_b)+\kappa_b\kappa_c(2\gamma_m\kappa_b+2\gamma_m\kappa_c+\kappa_b\kappa_c), \\		a_5=&-\Delta_b(4G^2\omega_m\kappa_c+g^2_c\Delta_c\gamma_m)+\gamma_m(\Delta_b^2+\kappa_b^2)(\Delta_c^2+\kappa_c^2)
+2\omega_m^2\Bigl[\Delta_c^2\kappa_b+\Delta_b^2\kappa_c+\kappa_c\kappa_b(\kappa_b+\kappa_c)\Bigr],\\ 		a_6=&-\Delta_b\omega_m\Bigl[g_c^2\Delta_c\omega_m+2G^2(\Delta_c^2+\kappa_c^2)\Bigr]
+\omega_m^2(\Delta_b^2+\kappa_b^2)(\Delta_c^2+\kappa_c^2).
	\end{aligned}
\end{equation}
\end{widetext}
Now we have $6$ Hurwitz determinants: $D_i$ with $i$ running from $1$ to $6$. The first two determinants share the same formation as Eq.~(\ref{determinant}). And the last four determinants read
\begin{equation}\label{determinants}
    \begin{aligned}
    D_3&=\begin{vmatrix}a_1&a_3&a_5\\a_0&a_2&a_4\\0&a_1&a_3\end{vmatrix},\quad D_4=\begin{vmatrix}a_1&a_3&a_5&0\\a_0&a_2&a_4&a_6\\0&a_1&a_3&a_5\\0&a_0&a_2&a_4\end{vmatrix}\\
D_5&=\begin{vmatrix}a_1&a_3&a_5&0&0\\a_0&a_2&a_4&a_6&0\\0&a_1&a_3&a_5&0
\\0&a_0&a_2&a_4&a_6\\0&0&a_1&a_3&a_5\end{vmatrix}, \quad D_6=D_5a_6.
    \end{aligned}
\end{equation}
Note again that $D_1$ is positive. Then all the other determinants should keep positive to meet the Routh-Hurwitz criterion. With a typical set of parameters that used in Fig.~\ref{NMS2}, $\kappa_b=10^{-2}\omega_m$, $\gamma_m=1.2\times10^{-3}\omega_m$, the blue detuning condition $\Delta_b=\Delta_c=-\omega_m$, and the destructive-interference condition by CQNC in Eq.~(\ref{MatchingCondition}), one can have an approximated stable condition for the linear system under control:
\begin{equation}
	|\alpha|^2\lesssim\frac{\omega_m^2}{9g^2}.
\end{equation}
Then for the system under CQNC, the upper-bound of the enhanced coupling strength could be estimated as one third of the mechanical frequency $G\lesssim\omega_m/3$.

Note following the similar procedure, the upper-bound of $G$ can also be evaluated for the imbalanced situation with $g_1\neq g_2$.

\bibliographystyle{apsrevlong}
\bibliography{ref}
\end{document}